\documentclass[12pt]{article}
\usepackage{a4wide}
\usepackage[T2A]{fontenc}
\usepackage[cp1251]{inputenc}
\usepackage[russian]{babel}
\usepackage{graphics}
\usepackage{amsmath, amsfonts, amssymb, amsthm,amscd}

\newcommand{\eps}{\varepsilon}
\newcommand{\ls}{\leqslant}
\newcommand{\gs}{\geqslant}

\newtheorem{state}{Утверждение}

\newtheorem{thm}{Теорема}
\newtheorem{dfn}{Определение}

\begin{document}

\title{О взаимосвязи мер кластеров и распределений расстояний в компактных метрических пространствах}
\author{Пушняков~A.\,C. \\
        \small{141700, Московская облаcть, г. Долгопрудный, Институтский пер., 9, МФТИ} \\
        \small{pushnyakovalex@mail.ru} \\
        \small{Поступила в редакцию 09.06.2016 г.}}
\date{}

\maketitle

\begin{abstract}

Рассматривается компактное метрическое пространство с ограниченной борелевской мерой. Под $r$"~кластером понимается любое измеримое множество диаметра не более $r$. Исследуется существование набора фиксированного числа $2r$"~кластеров обладающего следующими свойствами: кластеры попарно отделены друг от друга на расстояние $r$ и мера набора~--- суммарная мера кластеров набора~--- \textit{близка} к мере всего пространства. Показано, что среди таких наборов существует набор максимальной меры. Для распределения расстояний вводится $r$"~параметрическая дискретизация на \textit{короткие}, \textit{средние} и \textit{длинные} расстояния. В терминах данной дискретизации получена нижняя оценка на меру набора максимальной меры.

\textit{Ключевые слова}: кластеризация, компактное метрическое пространство, борелевская мера, метрика Хаусдорфа, теорема Бляшке, максимальное паросочетание.

\end{abstract}

\section{Введение}

Во многих задачах интеллектуального анализа данных для описания объектов используется метрическая информация \cite{zhur1971alg, aizer1970metod}. Для задач классификации и кластеризации \cite{celebi2013comparative, de2012minkowski, aggarwal2013data} предполагается, что используемая метрика удовлетворяет так называемому \textit{принципу компактности}: близкие объекты скорее должны лежать в одном классе, нежели в разных \cite{zagor1998, braverman1962}. В случае \textit{хорошей} метрики можно полагать, что множество объектов распадается на несколько кластеров, отделенных друг от друга.

Для метрических пространств, представимых в виде объединения кластеров, распределение расстояний имеет некоторые характерные особенности. Так, если у метрики есть некоторое характерное внутрикластерное расстояние $r$ и межкластерное расстояние $R$, причем $R  > r$, то можно ожидать, что доля расстояний в промежутке $(r, R)$ будет \textit{мала}. В данной статье исследуется следующий вопрос: что нужно потребовать от распределения расстояний, чтобы гарантировать наличие кластерной структуры в метрическом пространстве?

Мы рассматриваем компактные метрические пространства с ограниченной борелевской мерой (или, что тоже самое, компактные метрические тройки Громова \cite{gromov2007metric, vershik1998univer}). В данных терминах удобно определить кластерную структуру как набор фиксированного числа отделенных друг от друга кластеров. Ниже будет показано, что среди таких структур найдется структура максимальной меры. Тогда факт \textit{близости} меры данной структуры к мере всего пространства можно интерпретировать как представление метрики в виде объединения кластеров.

Наблюдение о внутрикластерных и межкластерных расстояниях может быть формализовано следующим образом: все расстояния разделяются на \textit{короткие}, \textit{средние} и \textit{длинные} и мы требуем, чтобы доля \textit{средних} расстояний была мала. Также мы потребуем выполнения некоторых дополнительных ограничений на распределение расстояний, обусловленных количеством кластеров в искомой кластерной структуре. В терминах параметрических ограничений на распределение расстояний мы получим нижнюю оценку на меру кластерной структуры максимальной меры. Вначале мы докажем искомую оценку для конечных полуметрических пространств с равномерной мерой, а затем, используя теорему Бляшке \cite{polovin2004}, обобщим оценку на случай компактного пространства.

\section{Постановка задачи}

Пусть дано компактное метрическое пространство~$(X, \rho)$ и ограниченная борелевская мера~$\mu$ на~$X$.
Любое борелевское подмножество~$X$ диаметра не более~$r$ будем называть~\textit{$r$-кластером}.
\begin{dfn}
Семейство $2r$"~кластеров $\mathcal{X} = \{X_1, \dots X_k\}$ будем называть $r$"~кластерной структурой порядка~$k$, если $\rho(X_i, X_j) \gs r$ при всех~$1 \ls i < j \ls k$, где~$\rho(A, B) = \inf \{\rho(x, y)\colon x \in A,\, y \in B\}$.
Мерой~$\mathcal{X}$ назовем величину $\mu(\mathcal{X}) \stackrel{\textrm{def}}{=} \displaystyle\sum_{i = 1}^k \mu(X_i)$.
\label{def:cluster_structure}
\end{dfn}

Верно следующее
\begin{state}
Среди всех $r$"~кластерных структур порядка~$k$ есть структура максимальной меры.
\label{state:max_cluster_existance}
\end{state}
\begin{proof}
Достаточно рассматривать структуры, содержащие только компактные множества.
Для каждой структуры произвольно занумеруем входящие в неё множества.
Пусть
$$
\mu^{*} = \sup_{\mathcal{X}}\mu(\mathcal{X}) < \infty
$$

Тогда найдется последовательность $\{\mathcal{X}_n\}^{\infty}_{n = 1}$ такая, что $\displaystyle\lim_{n \to \infty} \mu(\mathcal{X}_n) = \mu^{*}$.
По теореме Бляшке метрическое пространство компактов из~$X$ по метрике Хаусдорфа~$\rho_H$ является компактным.
Тогда без ограничения общности можно считать $X_{in} \stackrel{\rho_H}{\to} X_{i}^{*}$, где $\mathcal{X}_n = \{X_{in}\}_{i = 1}^k$.
Очевидно, что $\mathcal{X}^{*} = \{X_{i}^{*}\}_{i = 1}^k$ является $r$"~кластерной структурой порядка~$k$.
Пусть $\mu^* - \mu(\mathcal{X}) = \delta > 0$.
Рассмотрим множества $X_{i}^{\eps} = \{y \colon \rho(y, X_i^{*}) \ls \eps\}$. При достаточно малых~$\eps$
$$
\sum_{i = 1}^{k} (\mu(X_{i}^{\eps}) - \mu(X_{i}^{*})) \ls \frac{\delta}{2}, \;
\mu^* - \sum_{i = 1}^{k} \mu(X_{i}^{\eps}) > \frac{\delta}{2}.
$$
Но в силу сходимости $X_{in} \stackrel{\rho_H}{\to} X_{i}^{*}$ при фиксированном~$\eps$ и достаточно больших~$n$ выполнено $X_{in} \subset X_{i}^{\eps}$, и
$$
\mu(\mathcal{X}_n) = \sum_{i = 1}^k \mu(X_{in}) \ls \sum_{i = 1}^k \mu(X_{i}^{\eps}) < \mu^* - \frac{\delta}{2}.
$$
Получили противоречие.
\end{proof}

Нашей дальнейшей задачей является определение условий, при которых можно гарантировать, что отношение $\dfrac{\mu(\mathcal{X^{*}})}{\mu(X)}$ близко к единице, где $\mathcal{X^{*}}$~--- $r$-кластерная структура порядка~$k$ максимальной меры. Везде далее мы считаем~$k~\gs~2$.

Рассмотрим модельный пример: пусть $X = X_1 \sqcup \ldots \sqcup X_k$, где все множества~$X_i$ являются $r$-кластерами, и $\rho(X_i, X_j) \gs R > r$ при всех $1 \ls i < j \ls k$.
В данном случае мера $\frac{r}{2}$"~кластерной структуры порядка~$k$ максимальной меры равна~$\mu(X)$  и выполнены равенства:
$$
\{(x_1, \dots x_{k + 1}) \in X^{k + 1} \colon \rho(x_i, x_j) > r,\, 1 \ls i < j \ls k + 1\} = \varnothing
$$
$$
\{(x, y) \in X^2 \colon r < \rho(x, y) < R\} = \varnothing
$$

Пару точек $(x, y) \in X^2$ будем называть \textit{ребром}, длина ребра~--- это~$\rho(x,y)$ (более наглядная аналогия будет видна в случае~$|X| < \infty$).
В выше описанном примере нет ребер длины которых лежат в интервале $(r, R)$, а также среди любых~$k + 1$ точек есть ребро длины не больше~$r$.

Если $\rho(x, y) \ls r$, то будем называть ребро~$(x, y)$ \textit{$r$"~коротким}; если $\rho(x, y) > 3r$, то будем называть ребро~$(x, y)$ \textit{$r$"~длинным}; все остальные ребра~--- \textit{$r$"~средние}.
Набор точек $(x_1, \dots, x_{k})$ назовем \textit{$r$"~антикликой порядка $k$}, если $\rho(x_i, x_j) > r$ при всех $1 \ls i < j \ls k$.
Если понятно, о каком~$r$ идет речь, то приставка~$r$ будет опускаться.

В нижеследующих неравенствах~(\ref{eq:med_dist}) и~(\ref{eq:forbidden_clique}) пары $(x, y)$ и наборы $(x_1, \dots, x_{k + 1})$ мы считаем упорядоченными.
Потребуем, чтобы в нашем метрическом пространстве $(X, \rho)$ мера $r$"~средних ребер была \textit{мала} в следующем смысле:
\begin{equation}
M(X) = \frac{1}{2}\mu\{(x, y) \in X^2 \colon r < \rho(x, y) \ls 3r\} \ls \frac{1}{2}\alpha \mu(X)^2,
\label{eq:med_dist}
\end{equation}
а мера $r$"~антиклик порядка $k + 1$ была \textit{мала} в следующем смысле:
\begin{equation}
T_{k + 1}(X) = \frac{1}{(k + 1)!}\mu\{(x_1, \dots x_{k + 1}) \in X^{k + 1} \colon \rho(x_i, x_j) > r,\, 1 \ls i < j \ls k + 1\} \ls \frac{1}{(k + 1)!}\beta \mu(X)^{k + 1},
\label{eq:forbidden_clique}
\end{equation}
где $\alpha, \beta > 0$~--- параметры (мы будем их считать достаточно малыми).

Далее мы докажем, что при выполнении условий~(\ref{eq:med_dist}) и~(\ref{eq:forbidden_clique}) верна оценка меры $\mathcal{X^{*}}$ вида
\begin{equation}
\mu(\mathcal{X^{*}}) \gs \Psi(\alpha, \beta)\mu(X),
\label{eq:main_result}
\end{equation}
где~$\Psi(\alpha, \beta) \to 1$ при~$\alpha \to 0$ и~$\beta \to 0$.

Выбор верхней границы для интервала средних ребер объясняется техническими соображениями, и, гипотетически, может быть уменьшен.
Нижняя же граница увеличена быть не может, если мы хотим получить оценку вида~(\ref{eq:main_result}).
Это следует из следующего утверждения.
\begin{state}
Для любых~$\delta > 0$ и~$\alpha > 0$ существует компактное метрическое пространство $(X, \rho)$ такое, что
$$
\mu\{(x, y) \in X^2 \colon r < r' < \rho(x, y)\} \ls \alpha\mu(X)^2,
$$
и мера любого~$2r$ кластера не более $\delta \mu(X)$.
\label{state:lower_bound_med_dist}
\end{state}
\begin{proof}
Пусть $X = Y_1 \sqcup \ldots \sqcup Y_s$, и $|Y_i| = m$ при $1 \ls i \ls s$, а мера любого множества равна его мощности. Определим расстояние $\rho$ следующим образом:
$$
\rho(x, y) = \left\{
  \begin{array}{ll}
    2r', & x, y \in Y_i \\
    r', & x \in Y_i, y \in Y_j,\, i \neq j
  \end{array}
\right.
$$
Тогда мера любого $2r$"~кластера не превосходит~$s$, а
$$
\mu\{(x, y) \in X^2 \colon r < r' < \rho(x, y)\} = sm(m - 1)
$$
Осталось взять~$m > \frac{1}{\delta}$, $s > \frac{1}{\alpha}$.
\end{proof}

\section{Жадная кластерная структура}

Вначале мы получим оценку вида~(\ref{eq:main_result}) в случае, когда $(X, \rho)$~--- конечное полуметрическое пространство.
Отметим, что достаточно получить нижнюю оценку меры какой-то $r$"~кластерной структуры порядка $k$.
Рассмотрим следующую жадную процедуру.
Пусть $X_1$~--- множество максимальной мощности среди всех $2r$"~кластеров (если таких множеств несколько, то выберем любое). Обозначим его окрестность $r$ за $Z_1$, т.е.
$$
Z_1 = \{x \in X \colon \rho(x, X_1) < r\}
$$
Пусть у нас есть попарно непересекающиеся множества $Z_1, \dots, Z_m$. Тогда $X_{m + 1}$~--- множество максимальной мощности среди всех $2r$"~кластеров в $X \setminus \displaystyle\bigcup_{i = 1}^{m} Z_i$, а множество $Z_{m + 1}$~--- $r$"~окрестность $X_{m + 1}$ во множестве $X \setminus \displaystyle\bigcup_{i = 1}^{m} Z_i$, т.е.
$$
Z_{m + 1} = \left\{x \in X \setminus \bigcup_{i = 1}^{m} Z_i \colon \rho(x, X_{m + 1}) < r\right\}
$$
Так как мощность $X$ конечна, то процедура оборвётся не некотором шаге.

\begin{dfn}
Построенное разбиение $X = \displaystyle\bigsqcup_{i = 1}^n Z_i$ мы назовем жадным кластерным разбиением, а семейство $2r$"~кластеров $\{X_1,\dots X_k\}$ назовем жадной $r$"~кластерной структурой порядка~$k$.
\end{dfn}

Сделаем несколько замечаний относительно последнего определения. Во-первых, последовательности $Z_i$ и $X_i$ определяются неоднозначно~--- далее считается, что фиксирована некоторая пара последовательностей $(X_i, Z_i)$. Во-вторых, из построения очевидно, что жадная $r$"~кластерная структура порядка~$k$ является $r$"~кластерной структурой порядка~$k$ по определению~(\ref{def:cluster_structure}).

Отметим, что последовательность $\{|X|_i\}_{i = 1}^n$ монотонно убывает, однако, для последовательности $\{|Z|_i\}_{i = 1}^n$ свойство монотонности в общем случае не выполняется.
Пусть $\{W_i\}_{i = 1}^n$~--- упорядоченные по убыванию~$|Z_i|$.
Следующим шагом мы покажем, что в условиях~(\ref{eq:med_dist}) и~(\ref{eq:forbidden_clique}) и при достаточно малых $\alpha$ и $\beta$ первые $k$ по мощности~$Z_i$ покрывают \textit{почти все} множество~$X$, т.е. верно неравенство
\begin{equation*}
\sum_{i = 1}^k W_i \gs \Phi(\alpha, \beta)|X|,
\label{eq:external_conc}
\end{equation*}
где~$\Phi(\alpha, \beta) \to 1$ при~$\alpha \to 0$ и~$\beta \to 0$.

\section{Нижняя оценка числа антиклик}

Пусть $T_s(i_1, \dots, i_s)$~--- число $r$"~антиклик порядка $s$, таких, что ровно по одной вершине содержится в каждом из множеств $Z_{i_j}$.
Понятно, что при $s = 1$ выполнено $T_s(i_j) = |Z_{i_j}|$.
Нам понадобится следующее рекуррентное соотношение на $T_s(i_1, \dots, i_s)$.
\begin{state}
Пусть $i_1 < \ldots < i_s$, тогда при $s \gs 2$
\begin{equation*}
T_s(i_1, \dots, i_s) \gs \frac{|Z_{i_1}|}{s}T_{s - 1}(i_2, \dots, i_s),
\label{eq:clique_count_rec}
\end{equation*}
\label{state:clique_count_rec}
\end{state}
\begin{proof}
Пусть $x_2, \dots, x_s$~--- вершины некоторой антиклики, $x_j \in Z_{i_j}$. \
Для каждой из вершин~$x_{j}$ рассмотрим множества
$$
S(x_{j}) = \{y \in Z_{i_1}\colon \rho(x_{j}, y) \ls r\},
$$
Так как диаметр $S(x_j)$ не более $2r$, то $|S(x_j)| \ls |X_{i_1}|$.
Пусть $Y = Z_{i_1} \setminus \displaystyle\bigcup_{j = 2}^s S(x_j)$, тогда $|Y| \gs \dfrac{|Z_{i_1}|}{s}$.

Для любой точки $y \in Y$ вершины $y, x_2, \dots, x_s$ образуют $r$"~антиклику порядка $s$. Тогда имеем
$$
T_s(i_1, \dots, i_s) \gs \sum_{(x_2, \dots, x_s)} \frac{|Z_{i_1}|}{s} = \frac{|Z_{i_1}|}{s}T_{s - 1}(i_2, \dots, i_s)
$$
\end{proof}

Из утверждения~(\ref{state:clique_count_rec}) сразу же получаем
\begin{equation}
T_s(i_1, \dots, i_s) \gs \frac{1}{s!}\prod_{j = 1}^{s} |Z_{i_j}|
\label{eq:clique_count}
\end{equation}

Теперь мы получим нижнюю оценку на $T_{k + 1}(X)$~--- число $r$"~антиклик порядка $k~+~1$. Нам осталось только просуммировать неравенство~(\ref{eq:clique_count}) по всем наборам из~$k + 1$ множеств $Z_i$.
Введем обозначение для симметрического многочлена от $n$ переменных
\begin{equation*}
\sigma_s(y_1,\dots y_n) \stackrel{\textrm{def}}{=} \sum_{1 \ls i_1 < \ldots < i_s  \ls n}\prod_{j = 1}^{s}y_j,
\label{eq:symmetric_poly}
\end{equation*}
тогда, используя~(\ref{eq:forbidden_clique}) и~(\ref{eq:clique_count}), получим
\begin{equation*}
\frac{1}{(k + 1)!}\sigma_{k + 1}(z_1, \dots, z_n) \ls T_{k + 1}(X) \ls \frac{1}{(k + 1)!}\beta|X|^{k + 1}
\label{eq:clique_symm_bound}
\end{equation*}
Разделим каждое $z_i$ на $|X|$ и упорядочим по убыванию: получим набор $w_1 \gs \ldots \gs w_n$, и тогда
\begin{equation*}
\sigma_{k + 1}(w_1, \dots, w_n) \ls \beta
\label{eq:clique_symm_bound_norm}
\end{equation*}

\section{Нижняя оценка для $\sum_{j = 1}^k W_j$}

По сути мы получили следующую задачу оптимизации
\begin{equation}
    \left\{
    \begin{array}{l}
        f(\mathbf{w}) = \displaystyle\sum_{j = 1}^k w_j \to \min_{\mathbf{w}}\\
        w_i \geq 0\\
        w_i \geq w_j,\; i \leq j\\
        \displaystyle\sum_{i = 1}^n w_i = 1\\
        \sigma_{k + 1}(w_1, \dots, w_n) \ls c
    \end{array}
    \right.
\label{prob:opt1}
\end{equation}

Очевидно, что задача~(\ref{prob:opt1}) имеет решение.
Мы будем далее считать, что $n > k$, иначе решение задачи~(\ref{prob:opt1}) очевидно.
Нам понадобятся следующие простые утверждения.
\begin{state}
Пусть $\mathbf{w}$~--- решение задачи~(\ref{prob:opt1}), тогда либо~$w_i = w_j$ при всех $1 \ls i \ls j \ls n$, либо $\sigma_{k + 1}(w_1, \dots, w_n)~=~c$.
\label{state:opt_bound}
\end{state}
\begin{proof}
Предположим противное. Пусть $w_l > w_{l + 1}$.
Рассмотрим вектор
$$
\mathbf{w}^{\eps} = (w_1^{\eps}, \dots w_n^{\eps}) = (w_1 - (n - l)\eps, \dots, w_{l} - (n - l)\eps, w_{l + 1} + l\eps, \dots, w_n + l\eps)
$$
Так как $\sigma_{k + 1}(w_1, \dots, w_n) < c$, то при достаточно малых $\eps > 0$ вектор $\mathbf{w}^{\eps}$ будет допустимым для задачи~(\ref{prob:opt1}). Но $f(\mathbf{w}^{\eps}) < f(\mathbf{w})$, получили противоречие.
\end{proof}

\begin{state}
Пусть $\mathbf{w}$~--- решение задачи~(\ref{prob:opt1}) и $w_k = \lambda > 0$. Тогда $\mathbf{w} = (w_1, \underbrace{\lambda, \dots, \lambda}_{s}, \mu, 0, \dots, 0)$, где~$s \gs k - 1$ и $\mu~<~\lambda$.
\label{state:opt_cases}
\end{state}
\begin{proof}
Если все $w_i$ попарно равны $\lambda$, то утверждение верно.
Для любых $1 \ls i < j \ls n$ имеем
$$
\sigma_{k + 1}(w_1, \dots, w_n) = w_iw_j\sigma_{k - 1}(w_1, \dots, \hat{w_i}, \dots, \hat{w_j}, \dots ,w_n) +
$$
$$
+(w_j + w_i)\sigma_{k}(w_1, \dots, \hat{w_i}, \dots, \hat{w_j}, \dots ,w_n) + \sigma_{k + 1}(w_1, \dots, \hat{w_i}, \dots, \hat{w_j}, \dots ,w_n)
$$
Пусть нашлось $2 \ls i \ls k$ такое, что $w_i > \lambda = w_{i + 1} = \ldots = w_{k}$. Тогда рассмотрим вектор
$$
\mathbf{w}' = (w_1 + w_i - \lambda, w_2, \dots, w_{i - 1}, \lambda, \dots ,\lambda)
$$
$$
\sigma_{k + 1}(\mathbf{w}) - \sigma_{k + 1}(\mathbf{w}') = (w_1w_i - (w_1 + w_i - \lambda)\lambda)\sigma_{k - 1}(w_1, \dots, \hat{w_i}, \dots, \hat{w_j}, \dots ,w_n) > 0
$$
Пусть $w_l$~--- последняя ненулевая компонента $\mathbf{w}$. Пусть $j$~--- первая компонента $\mathbf{w}$ такая, что $w_j < \lambda$, и $j < l$. Тогда рассмотрим вектор
$$
\mathbf{w}' = (w_1, \dots w_{j - 1}, \min\{\lambda, w_l + w_j\}, w_{j + 1}, \dots, w_{l - 1}, \max\{w_l + w_j - \lambda, 0\}, \dots, 0)
$$
$$
\sigma_{k + 1}(\mathbf{w}) - \sigma_{k + 1}(\mathbf{w}') = (w_jw_l - \min\{\lambda, w_l + w_j\}\max\{w_l + w_j - \lambda, 0\})\sigma_{k - 1}(w_1, \dots, \hat{w_i}, \dots, \hat{w_j}, \dots ,w_n) > 0
$$
Тогда по утверждению~(\ref{state:opt_bound}) получаем, что $w$~--- не решение задачи~(\ref{prob:opt1}).
\end{proof}

Итак, рассмотрим вектор $\mathbf{w} = (w_1, \underbrace{\lambda, \dots, \lambda}_{s}, \mu, 0, \dots, 0)$.
$$
\sigma_{k + 1}(\mathbf{w}) = w_1\mu\sigma_{k - 1}(\underbrace{\lambda, \dots, \lambda}_{s}) + (w_1 + \mu)\sigma_{k}(\underbrace{\lambda, \dots, \lambda}_{s}) + \sigma_{k + 1}(\underbrace{\lambda, \dots, \lambda}_{s}) =
$$
\begin{equation}
= w_1\mu{s \choose k-1}\lambda^{k - 1} + (w_1 + \mu){s \choose k}\lambda^{k} + {s \choose k + 1}\lambda^{k + 1} \ls c
\label{eq:opt_bound}
\end{equation}

Мы рассмотрим несколько случаев.
\begin{enumerate}
\item $s = k - 1$. Неравенство~(\ref{eq:opt_bound}) переходит в
$$
\frac{1}{k + 1}\mu\lambda^{k - 1} \ls  w_1\mu\lambda^{k - 1} \ls c
$$
$$
f(\mathbf{w}) = 1 - \mu \gs 1 - \min\left\{\lambda, \frac{c(k + 1)}{\lambda^{k - 1}}\right\} \gs 1 - (c(k + 1))^{\frac{1}{k}}
$$

\item $s = k$.
$$
\frac{1}{k + 1}\lambda^k \ls w_1\mu k\lambda^{k - 1} + (w_1 + \mu)\lambda^k \ls c,\;\; \lambda \ls (c(k + 1))^{\frac{1}{k}}
$$
$$
f(\mathbf{w}) \gs 1 - 2\lambda \gs 1 - 2(c(k + 1))^{\frac{1}{k}}
$$

\item $s \gs k + 1$
$$
\left(\frac{s\lambda}{k + 1}\right)^{k + 1} \ls {s \choose k + 1}\lambda^{k + 1} \ls c,\;\; s\lambda \ls (k + 1)c^{\frac{1}{k + 1}}
$$
$$
f(\mathbf{w}) \gs 1 - s\lambda \gs 1 - (k + 1)c^{\frac{1}{k + 1}}
$$

\end{enumerate}

Так как при $k \gs 2$ и $c \ls \frac{1}{2}$ выполнено $(k + 1)c^{\frac{1}{k + 1}} \gs 2(c(k + 1))^{\frac{1}{k}}$, то верно следующее
\begin{state}
Пусть $\mathbf{w}$~--- решение задачи~(\ref{prob:opt1}) и $(k + 1)c^{\frac{1}{k + 1}} \ls 1$. Тогда
\label{sate:opt_solution}
\begin{equation*}
f(\mathbf{w}) \gs 1 - (k + 1)c^{\frac{1}{k + 1}}.
\label{eq:opt_solution}
\end{equation*}
\end{state}

Используя соотношения~(\ref{eq:clique_symm_bound_norm}) и~(\ref{eq:opt_solution}) получаем
\begin{equation*}
\sum_{i = 1}^k W_i \gs |X|\left(1 - (k + 1)\beta^{\frac{1}{k + 1}}\right)\stackrel{\mathrm{def}}{=}\Phi(\alpha, \beta)|X|.
\label{eq:external_conc_complete}
\end{equation*}

\section{Оценка меры жадной $r$-кластерной структуры}

Далее мы будем рассматривать только внутреннюю структуру множеств $Z_i$. Поэтому без ограничения общности можно полагать, что $|Z_i| = W_i$.

Нам осталось доказать, что
$$
\sum_{i = 1}^k |Z_i| - \sum_{i = 1}^k |X_i| = o(1)|X|, \; \alpha + \beta \to 0
$$

Рассмотрим множества~$Z_i$ и~$X_i \subset Z_i$.
Для любых $x \in X_i$ и $z \in Z_i$ выполнено $\rho(x, z) \ls 3r$, поэтому концы всех длинных ребер лежат в $Z_i \setminus X_i$.
Рассмотрим во множестве $Z_i \setminus X_i$ максимальное паросочетание из длинных ребер, которое покрывает множество $W_i$.
Пусть $Y_i = Z_i \setminus (X_i \cup W_i)$, тогда $X_i \cup Y_i$ является $3r$"~кластером.
Докажем простое утверждение, связывающее мощность~$X_i$ и число средних ребер в $X_i \cup Y_i$.
\begin{state}
Пусть $(A, \rho)$~--- конечное полуметрическое пространство диаметра не более~$3r$, и множество~$B$ является $2r$"~кластером максимальной мощности. Тогда число средних ребер не менее $M(A) \gs \frac{1}{2}|A||A\setminus B|$.
\label{sate:max_cluster}
\end{state}
\begin{proof}
Пусть~$x_0$~--- точка, из которой выходит максимально число коротких ребер, а~$S$~--- замкнутый шар радиуса~$r$ с центром в $x_0$. Тогда
$$
M(A) \gs \frac{1}{2}|A|(|A| - |S|) \gs \frac{1}{2}|A|(|A| - |B|)
$$
\end{proof}

Также для любого ребра $(u_1, u_2)$ из паросочетания, покрывающего $W_i$, и точки $x \in X_i$ хотя бы одно из ребер $(x, u_j)$ является средним.
В купе с утверждением~\ref{sate:max_cluster} получаем следующее неравенство:
\begin{equation*}
M(Z_i) \gs \frac{1}{2}(|X_i| + |Y_i|)|Y_i| + \frac{1}{2}|W_i||X_i|
\label{eq:med_int}
\end{equation*}

Сейчас мы применим технику аналогичную той, что использовалась при оценке числа антиклик.
\begin{state}
Пусть $T_s(Z_i)$~--- число $r$"~антиклик порядка $s$ во множестве $Z_i$. Тогда при $s \gs 3$
$$
T_s(Z_i) \gs \frac{1}{s}(|Z_i| - (s - 1)|X_i|)_{+}T_{s - 1}(Z_i)
$$
\end{state}
\label{state:clique_count_rec_int}
\begin{proof}
Доказательство почти дословно совпадает с доказательством утверждения~\ref{state:clique_count_rec}.
Пусть $x_1, \dots x_{s - 1}$ образуют некоторую антиклику.
Для каждой из вершин~$x_{j}$ рассмотрим множества
$$
S(x_{j}) = \{y \in Z_i\colon \rho(x_{j}, y) \ls r\},
$$
Так как диаметр $S(x_j)$ не более $2r$, то $|S(x_j)| \ls |X_{i_1}|$.
Пусть $Y = Z_i \setminus \displaystyle\bigcup_{j = 2}^s S(x_j)$, тогда
$$
|Y| \gs (|Z_i| - (s - 1)|X_i|)_{+}.
$$
Для любой точки $y \in Y$ вершины $y, x_1, \dots, x_{s - 1}$ образуют антиклику порядка $s$. Осталось заметить что каждую антиклику порядка $s$ мы посчитали не более $s$ раз, тогда имеем
$$
T_s(Z_i) \gs \sum_{(x_1, \dots, x_{s - 1})} \frac{1}{s}(|Z_i| - (s - 1)|X_i|)_{+} = \frac{1}{s}(|Z_i| - (s - 1)|X_i|)_{+}T_{s - 1}(Z_i).
$$
\end{proof}

Из утверждения~\ref{state:clique_count_rec_int} и равенства $T_1(Z_i) = |Z_i|$ сразу следует неравенство
\begin{equation*}
T_{k + 1}(Z_i) \gs \frac{1}{(k + 1)!}\prod_{j = 1}^{k + 1}(|Z_i| - (j - 1)|X_i|)_{+}
\label{eq:clique_count_int}
\end{equation*}

Если $|X_i|(k + 1) \ls |Z_i|$, то
$$
T_{k + 1}(Z_i) \gs \left(\frac{|Z_i|}{k + 1}\right)^{k + 1}
$$

Пусть $I_1$~--- множество всех индексов $1 \ls i \ls k$ таких, что $|X_i|(k + 1) \ls |Z_i|$, тогда
$$
\left(\frac{\sum_{i \in I_1}|Z_i|}{k(k + 1)}\right)^{k + 1}  \ls \frac{1}{(k + 1)^{k + 1}}\sum_{i \in I_1}|Z_i|^{k + 1} \ls \sum_{i \in I_1} T_{k + 1}(Z_i) \ls T_{k + 1}(X) \ls \frac{\beta |X|^{k + 1}}{(k + 1)!}
$$
$$
\sum_{i \in I_1}|Z_i| \ls ek\beta^{\frac{1}{k + 1}} |X|
$$

Если же $|X_i|(k + 1) > |Z_i|$, то из неравенства~(\ref{eq:med_int}) получаем
$$
M(Z_i) \gs \frac{1}{2(k + 1)}|Z_i|(|Y_i| + |W_i|) = \frac{1}{2(k + 1)}|Z_i|(|Z_i| - |X_i|)
$$
$$
|Z_i| - |X_i| \ls \frac{2(k + 1)M(Z_i)}{|Z_i|}
$$

Рассмотрим $I_2$~--- множество таких индексов $1 \ls i \ls k$, что $i \notin I_1$ и $|Z_i| \gs \sqrt{\alpha} |X|$. Тогда суммируя предыдущее неравенство по множеству $I_2$:
$$
\sum_{i \in I_2}(|Z_i| - |X_i|) \ls \frac{2(k + 1)M(X)}{\sqrt{\alpha}|X|} \ls \sqrt{\alpha}(k + 1)|X|
$$
Наконец, получаем
$$
\sum_{i = 1}^k(|Z_i| - |X_i|) \ls \sum_{i \in I_1}|Z_i| + \sum_{i \in I_2}(|Z_i| - |X_i|) + \sum_{i \notin I_1\cup I_2}|Z_i| \ls
$$
$$
\ls ek\beta^{\frac{1}{k + 1}} |X| + \sqrt{\alpha}(k + 1)|X| + \sqrt{\alpha}k|X| = (\sqrt{\alpha}(2k + 1) + ke\beta^{\frac{1}{k + 1}})|X|
$$
Итак, мы доказали следующую теорему
\begin{thm}
Пусть $(X, \rho)$ конечное полуметрическое пространство с равномерной мерой $\mu$, а $\mathcal{X}^*$~--- $r$"~кластерная структура максимальной меры. Тогда, если выполнены неравенства~(\ref{eq:med_dist}) и~(\ref{eq:forbidden_clique}), то
\begin{equation}
\mu(\mathcal{X}^*) \gs \Psi(\alpha, \beta)|X|,
\label{eq:main_th}
\end{equation}
где
$$
\Psi(\alpha, \beta) = 1 - \sqrt{\alpha}(2k + 1) - (k(e + 1) + 1)\beta^{\frac{1}{k + 1}}
$$
\label{th:finite}
\end{thm}

\section{Обобщение на случай произвольного компактного пространства}

Мы будем использовать технику, аналогичную той, что была использована при доказательстве утверждения~(\ref{state:max_cluster_existance}).
\begin{thm}
Пусть $(X, \rho)$ компактное метрическое пространство с ограниченной борелевской мерой $\mu$, а $\mathcal{X}^*$~--- $r$"~кластерная структура максимальной меры. Тогда, если выполнены неравенства~(\ref{eq:med_dist}) и~(\ref{eq:forbidden_clique}), то выполнено неравенство~(\ref{eq:main_th}).
\end{thm}
\begin{proof}
Фиксируем произвольное $0 < \eps < 1$. В $X$ существует конечная~$\eps$"~сеть, а значит и разбиение~$X$ на конечное число~$N_{\eps}$ $\eps$"~кластеров $\{A_i\}_{i = 1}^{N_{\eps}}$.
Выберем $N_{\eps}$ положительных рациональных чисел $q_1, \dots q_{N_{\eps}}$ так, что $\mu(A_i) \gs q_i$ при $1 \ls i \ls N_{\eps}$ и $q_i \gs \mu(A_i)(1 - \eps)$.

Рассмотрим полуметрическое пространство конечной мощности $X_{\eps} = B_1 \sqcup \ldots \sqcup B_s$, где $\dfrac{|B_i|}{|B_j|} = \dfrac{q_i}{q_j}$, а функция расстояния $\rho_{\eps}$ определяется следующим образом:
$$
\rho_{\eps}(x, y) = \left\{
  \begin{array}{ll}
    0, & x, y \in B_i \\
    \rho(A_i, A_j), & x \in B_i, y \in B_j,\, i \neq j
  \end{array}
\right.
$$
Отметим, что
$$
|B_i| = \frac{q_i |X_{\eps}|}{\displaystyle\sum_{j = 1}^{N_{\eps}}q_j} \ls \frac{\mu(A_i) |X_{\eps}|}{(1 - \eps)\displaystyle\sum_{j = 1}^{N_{\eps}}\mu(A_j)} = \frac{\mu(A_i) |X_{\eps}|}{(1 - \eps)\mu(X)}
$$
Если $x \in B_i$, $y \in B_j$ и $\rho_{\eps}(x, y) \in (r, 3r]$, то для всех $v \in A_i$, $u \in A_j$ верно $\rho(v, u) \in (r, 3r + 2\eps]$. Отсюда получаем оценку на число $r$"~средних ребер в $X_{\eps}$:
$$
M(X_{\eps}) = \sum_{1 \ls i < j \ls N_{\eps}}[\rho(A_i, A_j) > r]|B_i||B_j| \ls
$$
$$
\ls \frac{|X_{\eps}|^2}{(1 - \eps)^2\mu(X)^2}\sum_{1 \ls i < j \ls N_{\eps}}[\rho(A_i, A_j) > r] \mu(A_i)\mu(A_i) \ls
$$
$$
\ls \frac{|X_{\eps}|^2}{2(1 - \eps)^2\mu(X)^2}\mu\{(x, y) \in X^2 \colon r < \rho(x, y) \ls 3r + 2\eps\} \stackrel{(\ref{eq:med_dist})}{\ls}
$$
$$
\ls \frac{|X_{\eps}|^2}{2(1 - \eps)^2}\left(\alpha + \frac{1}{\mu(X)^2}\mu\{(x, y) \in X^2 \colon 3r < \rho(x, y) \ls 3r + 2\eps\}\right) \stackrel{\textrm{def}}{=} \frac{1}{2}\alpha_{\eps}|X_{\eps}|^2
$$
Аналогично имеем оценку для $r$"~антиклик порядка~$k + 1$:
$$
T_{k + 1}(X_{\eps}) = \sum_{1 \ls i_1 < \dots < i_{k + 1} \ls N_{\eps}}\prod_{1 \ls j < l \ls k + 1} [\rho(A_{i_j}, A_{i_l}) > r] \prod_{j = 1}^{k + 1} |B_{i_j}| \ls
$$
$$
\ls \frac{|X_{\eps}|^{k + 1}}{(1 - \eps)^{k + 1}\mu(X)^{k + 1}}\sum_{1 \ls i_1 < \dots < i_{k + 1} \ls N_{\eps}}\prod_{1 \ls j < l \ls k + 1} [\rho(A_{i_j}, A_{i_l}) > r] \prod_{j = 1}^{k + 1} \mu(A_{i_j}) \stackrel{(\ref{eq:forbidden_clique})}{\ls}
$$
$$
\ls \frac{|X_{\eps}|^{k + 1}\beta}{(k + 1)!(1 - \eps)^{k + 1}} \stackrel{\textrm{def}}{=} \frac{1}{(k + 1)!} \beta_{\eps}|X_{\eps}|^{k + 1}
$$
Заметим, что при $\eps \to 0$ $\alpha_{\eps} \to \alpha$ и $\beta_{\eps} \to \beta$.
В силу теоремы~\ref{th:finite} получаем, что в~$X_{\eps}$ существует $r$"~кластерная структура порядка $k$ $\mathcal{C}_{\eps} = \{C_{1\eps}, \dots C_{k\eps}\}$ меры не менее $\Psi(\alpha_{\eps}, \beta_{\eps})|X|_{\eps}$.

Понятно, что каждое $B_i$ либо полностью содержится в каком-то множестве семейства $\mathcal{C}_{\eps}$, либо никакой элемент $B_i$ не входит ни в какое множество семейства $\mathcal{C}_{\eps}$.
Для каждого $C_{i\eps} = B_{j_1} \sqcup \ldots \sqcup B_{j_l}$ рассмотрим множество $X_{i\eps} = cl(A_{j_1} \sqcup \ldots \sqcup A_{j_l})$ в $X$. Заметим, что множество $X_{i\eps}$ является $(r + 2\eps)$"~кластером, и для любых $1 \ls i < j \ls k$ выполнено $\rho(X_{i\eps}, X_{j\eps}) \gs r$.

Настало время снова применить теорему Бляшке.
Пусть $\eps \to 0$.
Рассматривая последовательность наборов $(X_{1\eps},\ldots X_{k\eps})$, без ограничения общности можно считать, что $X_{i\eps} \stackrel{\rho_{H}}{\to} X_i^{*}$.
Очевидно, что множества $X_i^*$ образуют $r$"~кластерную структуру порядка $k$, назовем её $\mathcal{X^*}$
Более того
$$
\sum_{i = 1}^k \mu(X_{i\eps}) \gs \frac{(1 - \eps)\mu(X)}{|X_{\eps}|}\sum_{i = 1}^k |C_{i\eps}| \gs (1 - \eps)\Psi(\alpha_{\eps}, \beta_{\eps})\mu(X) \stackrel{\eps \to 0}{\to} \Psi(\alpha, \beta)\mu(X).
$$
Почти дословно повторяя доказательство утверждения~(\ref{state:max_cluster_existance}), получаем $\mu(\mathcal{X^*}) \gs \Psi(\alpha, \beta)\mu(X)$.

\end{proof}


\begin{thebibliography}{10}
\def\selectlanguageifdefined#1{
\expandafter\ifx\csname date#1\endcsname\relax
\else\language\csname l@#1\endcsname\fi}
\ifx\undefined\url\def\url#1{{\small #1}}\else\fi
\ifx\undefined\BibUrl\def\BibUrl#1{\url{#1}}\else\fi
\ifx\undefined\BibAnnote\long\def\BibAnnote#1{}\else\fi
\ifx\undefined\BibEmph\def\BibEmph#1{\emph{#1}}\else\fi

\bibitem{zhur1971alg}
\selectlanguageifdefined{russian}
\BibEmph{Журавлев~Ю.И., Никифоров~В.В.} Алгоритмы распознавания, основанные
  на вычислении оценок~// \BibEmph{Кибернетика}.
\newblock 1971.
\newblock {\cyr\textnumero}~3.
\newblock {\cyr\CYRS.}~1--11.

\bibitem{aizer1970metod}
\selectlanguageifdefined{russian}
\BibEmph{Айзерман~М.А., Браверман~Э.М., Розоноэр~Л.И.} Метод потенциальных
  функций в теории обучения машин.
\newblock М.: Наука, 1970.

\bibitem{celebi2013comparative}
\selectlanguageifdefined{english}
\BibEmph{Celebi~M.E., Kingravi~H.A., Vela~P.A.} A comparative study of
  efficient initialization methods for the k-means clustering algorithm~//
  \BibEmph{Expert Systems with Applications}.
\newblock 2013.
\newblock Vol.~40, no.~1.
\newblock Pp.~200--210.

\bibitem{de2012minkowski}
\selectlanguageifdefined{english}
\BibEmph{De~Amorim~R.C., Mirkin~B.} Minkowski metric, feature weighting and
  anomalous cluster initializing in k-means clustering~// \BibEmph{Pattern
  Recognition}.
\newblock 2012.
\newblock Vol.~45, no.~3.
\newblock Pp.~1061--1075.

\bibitem{aggarwal2013data}
\selectlanguageifdefined{english}
\BibEmph{Aggarwal~C.C., Reddy~C.K.} Data clustering: algorithms and
  applications.
\newblock CRC Press, 2013.

\bibitem{zagor1998}
\selectlanguageifdefined{russian}
\BibEmph{Загоруйко~Н.Г.} Гипотезы компактности и $\lambda$-компактности в
  методах анализа данных~// \BibEmph{Сибирский журнал индустриальной
  математики}.
\newblock 1998.
\newblock \CYRT.~1, {\cyr\textnumero}~1.
\newblock {\cyr\CYRS.}~114--126.

\bibitem{braverman1962}
\selectlanguageifdefined{russian}
\BibEmph{Браверман~Э.М.} Опыты по обучению машины распознаванию зрительных
  образов~// \BibEmph{Автоматика и телемеханика}.
\newblock 1962.
\newblock \CYRT.~23, {\cyr\textnumero}~3.
\newblock {\cyr\CYRS.}~349--365.

\bibitem{gromov2007metric}
\selectlanguageifdefined{english}
\BibEmph{Gromov~M.} Metric structures for Riemannian and non-Riemannian spaces.
\newblock Springer Science \& Business Media, 2007.

\bibitem{vershik1998univer}
\selectlanguageifdefined{russian}
\BibEmph{Вершик~А.М.} Универсальное пространство урысона, метрические тройки
  громова и случайные метрики на натуральном ряде~// \BibEmph{Успехи
  математических наук}.
\newblock 1998.
\newblock \CYRT.~53, {\cyr\textnumero} 5 (323).
\newblock {\cyr\CYRS.}~57--64.

\bibitem{polovin2004}
\selectlanguageifdefined{russian}
\BibEmph{Половинкин~E.C., Балашов~М.В.} Элементы выпуклого и сильно выпуклого
  анализа.
\newblock М.: Физматлит, 2004.

\end{thebibliography}

\end{document}